\documentclass[a4paper,aps,floatfix,twocolumn,superscriptaddress,10pt,accepted=2018-01-16]{quantumarticle}
\usepackage{graphicx}
\usepackage{amsmath}
\usepackage{amssymb}
\usepackage{bm}

\begin{document}

\title{Machine-learning-assisted correction of correlated qubit errors in a topological code}
\author{P. Baireuther}
\affiliation{Instituut-Lorentz, Universiteit Leiden, P.O. Box 9506, 2300 RA Leiden, The Netherlands}
\author{T. E. O'Brien}
\affiliation{Instituut-Lorentz, Universiteit Leiden, P.O. Box 9506, 2300 RA Leiden, The Netherlands}
\author{B. Tarasinski}
\affiliation{QuTech, Delft University of Technology, P.O. Box 5046, 2600 GA Delft, The Netherlands}
\author{C. W. J. Beenakker}
\affiliation{Instituut-Lorentz, Universiteit Leiden, P.O. Box 9506, 2300 RA Leiden, The Netherlands}

\date{December 2017}
\begin{abstract} 
A fault-tolerant quantum computation requires an efficient means to detect and correct errors that accumulate in encoded quantum information. In the context of machine learning, neural networks are a promising new approach to quantum error correction. Here we show that a recurrent neural network can be trained, \textit{using only experimentally accessible data}, to detect errors in a widely used topological code, the surface code, with a performance above that of the established minimum-weight perfect matching (or ``blossom'') decoder. The performance gain is achieved because the neural network decoder can detect correlations between bit-flip (X) and phase-flip (Z) errors. The machine learning algorithm adapts to the physical system, hence no noise model is needed. The long short-term memory layers of the recurrent neural network maintain their performance over a large number of quantum error correction cycles, making it a practical decoder for forthcoming experimental realizations of the surface code.
\end{abstract} 
\maketitle

\section{Introduction}
\label{sec_intro}
\begin{figure*}[t]
\centering{ 
\includegraphics[width=0.8\textwidth]{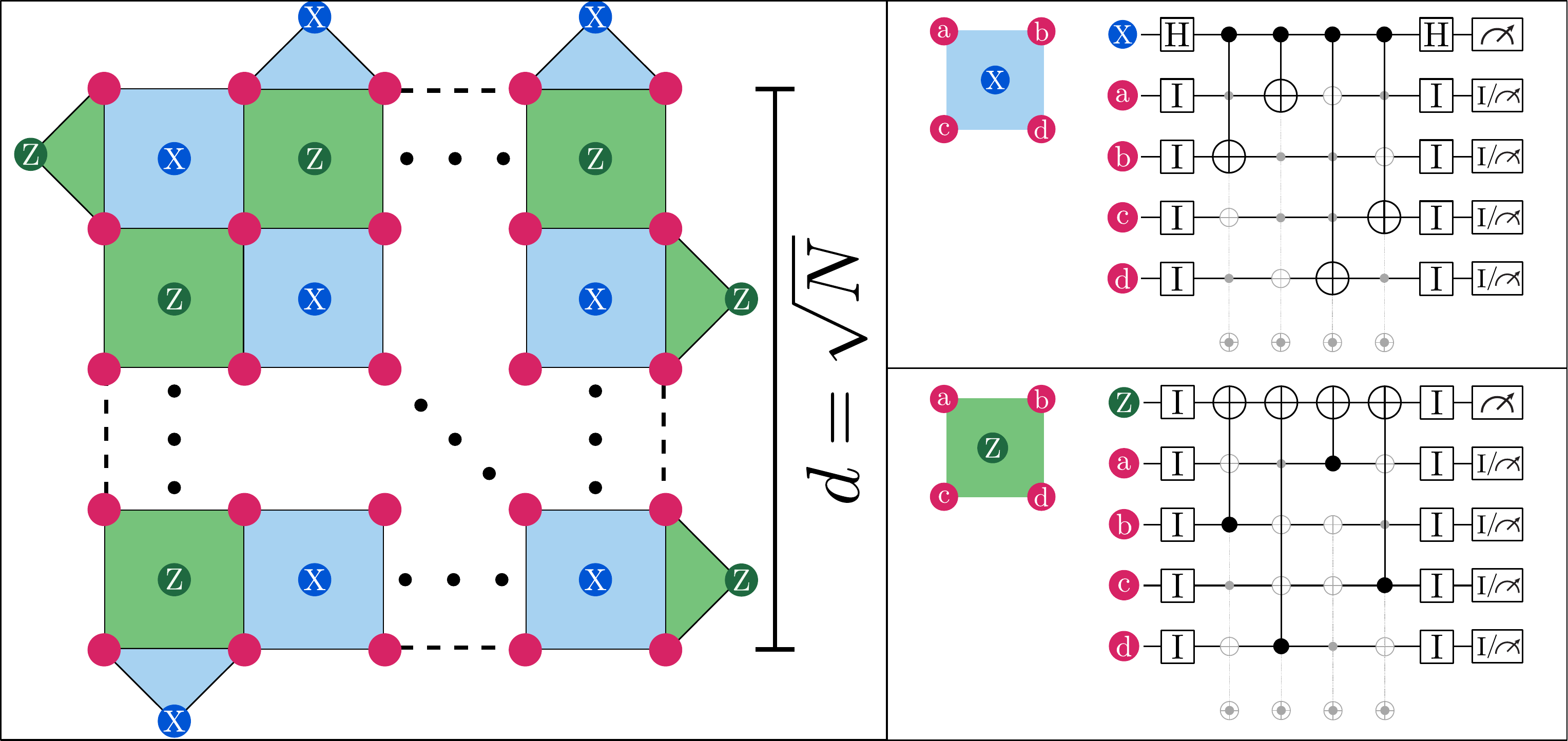}
\caption{\label{fig:Surface_code}Schematic of the surface code. \textit{Left:} $N$ physical data qubits are arranged on a $d\times d$ square lattice (where $d=\sqrt{N}$ is known as the distance of the code). For each square one makes the four-fold $\sigma_x$ or $\sigma_z$ correlated measurement of Eq.~\eqref{eq:stabilizers}. A further set of two-fold $\sigma_x$ and $\sigma_z$ measurements are performed on the boundary, bringing the total number of measurements to $N-1$. \textit{Right:} Since direct four-fold parity measurements are impractical, the measurements are instead performed by entanglement with an ancilla qubit, followed by a measurement of the ancilla in the computational basis. Both data qubits and ancilla qubits accumulate errors during idle periods (labeled I) and during gate operations (Hadamard H and {\sc cnot}), which must be accounted for by a decoder. The data qubits are also entangled with the rest of the surface code by the grayed out gates. 
}
}
\end{figure*}
A quantum computer needs the help of a powerful classical computer to overcome the inherent fragility of entangled qubits. By encoding the quantum information in a nonlocal way, local errors can be detected and corrected without destroying the entanglement~\cite{Lid13,Ter15}. Since the efficiency of the quantum error correction protocol can make the difference between failure and success of a quantum computation, there is a major push towards more and more efficient decoders~\cite{Fow12a}. Topological codes such as the surface code, which store a logical qubit in the topology of an array of physical qubits, are particularly attractive because they combine a favorable performance on small circuits with scalability to larger circuits~\cite{Bra98,Wan10,Fow12b,Tom14,Woo16,Nic16}.

In a pioneering work~\cite{Tor16}, Torlai and Melko have shown that the data processing power of machine learning (artificial neural networks \cite{Roj96,Ben09,Sha14}) can be harnessed to produce a flexible, adaptive decoding algorithm. A test on a topological code (Kitaev's toric code~\cite{Kit03}) revealed a performance for phase-flip errors that was comparable to decoders based on the minimum-weight perfect matching (MWPM or ``blossom'') algorithm of Edmonds~\cite{Edm65,Den01,Fow15}. The machine learning paradigm promises a flexibility that the classic algorithms lack, both with respect to different types of topological codes and with respect to different types of errors. 

Several groups are exploring the capabilities of a neural network decoder \cite{Var17,Kra17,note0}, but existing designs cannot yet be efficiently deployed as a decoder in a surface code architecture \cite{Kel15,Tak16,Ver16}. Two key features which are essential for this purpose are 1: The neural network must have a ``memory'', in order to be able to process repeated cycles of stabilizer measurement whilst detecting correlations between cycles; and 2: The network must be able to learn from measured data, it should not be dependent on the uncertainties of theoretical modeling.

In this work we design a recurrent neural network decoder that has both these features, and demonstrate a performance improvement over a blossom decoder in a realistic simulation of a forthcoming error correction experiment. Our decoder achieves this improvement through its ability to detect bit-flip (X) and phase-flip (Z) errors separately as well as correlations (Y). The blossom decoder treats a Y-error as a pair of uncorrelated X and Z errors, which explains the improved performance of the neural network. We study the performance of the decoder in a simplified model where the Y-error rate can be adjusted independently of the X- and Z-error rates, and measure the decoder efficiency in a realistic model (density matrix simulation) of a state-of-the-art 17-qubit surface code experiment (Surface-17).

The outline of this paper is as follows. In the next section \ref{sec_surfacecode} we summarize the results from the literature we need on quantum error correction with the surface code. The design principles of the recurrent neural network that we will use are presented in Sec.\ \ref{sec_correlatederrs}, with particular attention for the need of an internal memory in an efficient decoder. This is one key aspect that differentiates our recurrent network from the feedforward networks proposed independently \cite{Var17,Kra17,note0} (see Sec.\ \ref{sec_related_work}). A detailed description of the architecture and training protocol is given in Sec.\ \ref{sec_neuralnetwork}. In Sec.\ \ref{sec_performance} we compare the performance of the neural network decoder to the blossom decoder for a particular circuit model with varying error rates. We conclude in Sec.\ \ref{sec_conclusion} with a demonstration of the potential of machine learning for real-world quantum error correction, by decoding data from a realistic quantum simulation of the Surface-17 experiment.

\section{Overview of the surface code}
\label{sec_surfacecode}
To make this paper self-contained we first describe the operation of the surface code and formulate the decoding problem. The expert reader may skip directly to the next section.

In a quantum error correcting (QEC) code, single logical qubits (containing the quantum information to be protected) are spread across a larger array of $N$ noisy physical data qubits~\cite{Sho95,Ste96}.
The encoding is achieved by $N-1$ binary parity check measurements on the data qubits~\cite{Got97}. Before these measurements, the state of the physical system is described by a complex vector $|\psi\rangle$ within a $2^N$-dimensional Hilbert space $\mathcal{H}$.
Each parity check measurement $M_i$ projects $|\psi\rangle$ onto one of two $2^{N-1}$-dimensional subspaces, dependent on the outcome $s_i$ of the measurement.
As all parity check measurements commute, the result of a single cycle of $N-1$ measurements is to project $|\psi\rangle$ into the intersection of all subspaces $\mathcal{H}_{\vec{s}}$ decided by the measurements $\vec{s}={s_1,\ldots,s_{N-1}}$ ($s_i\in\{0,1\}$).
This is a Hilbert space of dimension $2^N/2^{N-1}=2$, giving the required logical qubit $|\psi\rangle_L$.

Repeated parity check measurements $\vec{s}(t)$ do not affect the qubit within this space, nor entanglement between the logical qubit states and other systems.
However, errors in the system will cause the qubit to drift out of the logical subspace.
This continuous drift is discretized by the projective measurement, becoming a series of discrete jumps between subspaces $\mathcal{H}_{\vec{s}(t)}$ as time $t$ progresses.
Since $\vec{s}(t)$ is directly measured, the qubit may be corrected, \textit{i.e.} brought back to the initial logical subspace $\mathcal{H}_{\vec{s}(0)}$.
When performing this correction, a decision must be made on whether to map the logical state $|0\rangle_L^{\vec{s}(t)}\in\mathcal{H}_{\vec{s}(t)}$ to $|0\rangle_L^{\vec{s}(0)}$ or $|1\rangle_L^{\vec{s}(0)}\in\mathcal{H}_{\vec{s}(0)}$, as no \textit{a priori} relationship exists between the labels in these two spaces.
If this is done incorrectly, the net action of the time evolution and correction is a logical bit-flip error.
A similar choice must be made for the $\{|+\rangle_L, |-\rangle_L\}$ logical states, which if incorrect results in a logical phase-flip error.

Information about the best choice of correction (to most-likely prevent logical bit-flip or phase-flip errors) is stored within the measurement vectors $\vec{s}$, which detail the path the system took in state-space from $\mathcal{H}_{\vec{s}(0)}$ to $\mathcal{H}_{\vec{s}(t)}$.
The non-trivial task of decoding, or extracting this correction, is performed by a classical decoder.
Optimal (maximum-likelihood) decoding is an NP-hard problem~\cite{Hsi10}, except in the presence of specific error models~\cite{Bra14}.
However, a fault-tolerant decoder need not be optimal, and polynomial time decoders exist with sufficient performance to demonstrate error mitigation on current quantum hardware~\cite{Wan10}.
This sub-optimality is quantified by the decoder efficiency~\cite{Obr17}
\begin{equation}
\eta_d=\epsilon_L^{\mathrm{(opt)}}/\epsilon_L^{\mathcal{D}}\label{eq:decoder_efficiency},
\end{equation}
where $\epsilon_L^{\mathcal{D}}$ is the probability of a logical error per cycle using the decoder $\mathcal{D}$, and $\epsilon_L^{\mathrm{(opt)}}$ is the probability of a logical error per cycle using the optimal decoder~\cite{Hei16}.

The QEC code currently holding the record for the best performance under a scalable decoder is the surface code~\cite{Bra98,Den01,Wan10,Fow12a}.
As illustrated in Fig.~\ref{fig:Surface_code}, the surface code is defined on a $d \times d$ lattice of data qubits, where $d=\sqrt{N}$ is the distance of the code.
The measurement operators are defined by coloring lattice squares as on a checkerboard.
Each square corresponds to a correlated measurement of the stabilizer operator
\begin{equation}
{\cal S}_{\alpha}=\sigma^{\text{a}}_{\alpha}\otimes\sigma^{\text{b}}_{\alpha}\otimes\sigma^{\text{c}}_{\alpha}\otimes\sigma^{\text{d}}_{\alpha},\label{eq:stabilizers}
\end{equation}
with $\alpha=z$ on the green squares and $\alpha=x$ on the blue squares.
The operator $\sigma^{\text{D}}_{\alpha}$ is the Pauli matrix acting on the qubit in the D-corner  of the square (labeled a,b,c,d in Fig~\ref{fig:Surface_code}).
The checkerboard is extended slightly beyond the boundary of the lattice~\cite{Bom07}, giving an additional set of two-qubit $\sigma_{\alpha}^{\text{D}}\sigma_{\alpha}^{\text{D}'}$ measurements, and bringing the total number of measurements to $(d-1)^2+2(d-1)=N-1$, as it should be.

All measurements commute because green and blue squares either share two corners or none.
A bit-flip or phase-flip on any data qubit in the bulk of the code causes two measurements to change sign, producing unit syndrome increments
\begin{equation}
\delta s_i(t)\equiv s_i(t)-s_i(t-1)\mod 2 .\label{eq:syndrome_derivative}
\end{equation}
This theme is continued even when the measurement of $s_i$ itself is allowed to be faulty; such measurement errors cause two correlated error signals $\delta s_i(t)=1$ separated in time, rather than in space.

As all observable errors can be built from combinations of bit-flip and phase-flip errors, these measurements allow the mapping of surface-code decoding to the minimum-weight perfect matching (MWPM) problem~\cite{Den01,Wan10}.
Every instance of non-zero $\delta s_i(t)$ is mapped to a vertex in a graph, with an edge between two vertices representing the probability of some combination of errors causing these signals. 
A `boundary' vertex is included to account for qubits on the edge of the lattice, whose errors may only cause a single error signal.
Then, the most probable matching of vertices, weighted by the product of probabilities on individual edges,
gives the required error correction.
This matching can be found in polynomial time with Edmonds' blossom algorithm~\cite{Edm65}.

Under current experimental parameters, with the smallest non-trivial $N$ ($N=9$, or distance $d=\sqrt{N}=3$), this blossom decoder already crosses the quantum memory threshold --- whereby quantum information on a logical qubit can be stored for a longer time than on any physical component.
However, the decoder itself performs only with efficiency $\eta_d=0.64$, leaving much room for improvement~\cite{Obr17}.

\section{Neural network detection of correlated errors}
\label{sec_correlatederrs}
The sub-optimality of the blossom decoder comes primarily from its inability to optimally detect Pauli-Y ($\sigma_y$) errors~\cite{Hei16,Obr17,Fow13}.
These errors correspond to a combination of a bit-flip (X) and a phase-flip (Z) on the same qubit, and are thus treated by a MWPM decoder as two independent errors.
Since these correlations exist as patterns on the graph, one may expect that the pattern matching capabilities of a neural network could be exploited to identify the correlations, producing an improvement over existing decoders.
This is the primary motivation of the research we report in what follows.

A key issue in the design of any practical decoder is to ensure that the decoder is able to operate for an \textit{unspecified} number of cycles $T$.
A feedforward neural network is trained on a dataset with a specific fixed $T$.
The central advance of this work is to use a \textit{recurrent} neural network to efficiently decode an arbitrary, unspecified number of cycles. 
In order to learn time correlations the network possesses an internal memory that it utilizes to store information about previous cycles. This is important because errors on the ancilla qubits or during ancilla qubit readout lead to error signals that are correlated over several cycles.

We adopt the recurrent neural network architecture known as a ``long short-term memory'' (LSTM) layer~\cite{Hoc97, Zar14}. (See App.\ \ref{details_appendix} for details of our network.)
These layers have two internal states: a short-term memory $\vec h_t$, and a long-term memory $\vec c_t$ that is updated each cycle and retains information over several cycles.
During training, the parameters that characterize the LSTM layers are updated using back propagation, in order to efficiently update and utilize the long-term memory to detect logical errors, even if the corresponding syndrome patterns are non-local in time. The parameters of the LSTM layers themselves are the same for each cycle; only their memory changes. This allows for a very efficient algorithm, whose computational cost per cycle is independent of how many cycles the network has to decode.

We now formulate the QEC problem that a decoder needs to solve.
To be useful for upcoming QEC experiments and future fault-tolerant quantum algorithms, it is critical that any decoder uses data that could be generated by such experiments.
This implies that the data available to the neural network, both for input and labels, must be data generated by qubit measurements (as opposed to a listing of occurred errors, which is not available in an actual experiment).

The data available to the decoder after $T$ cycles are the $T$ syndromes $\vec{s}(t)$, and a final syndrome $\vec{f}$ calculated from readout of the final data qubits.
From this, a decoder must output a single bit of data, the so-called ``final parity correction'' that decides on the correction of the final logical state. The decoder may be trained and tested using the scheme described in Ref.\ \onlinecite{Obr17}.
The system is prepared in a known logical state, chosen from $|0\rangle_L$ and $|1\rangle_L$ or from $|+\rangle_L$ and $|-\rangle_L$, which is held for $T$ cycles and then readout.
The final logical state can be determined by the parity of all data qubit measurements, to which the final parity correction may be directly added.
This gives a standard binary classification problem for the neural network.
Since it is a priori unknown in which basis the logical qubit will be measured, we need to train two separate decoders --- one for the $x$-basis and one for the $z$-basis. 

\section{Related Work}
\label{sec_related_work}

\subsection{Approaches going beyond blossom decoding}

The neural network decoder improves  on the blossom decoder by including correlations between Pauli-X and Pauli-Z errors. It is possible to account for these correlations without using machine learning, by adapting the minimum-weight perfect matching (blossom) algorithm. 

Fowler \cite{Fow13} and Delfosse and Tillich \cite{Del14} achieved this by performing repeated rounds of X-error and Z-error decoding in series.
After each round of X-error decoding, the weights on the Z-graph are updated based on the likelihood of underlying Z-errors assuming the X-matching is correct.
The overhead from repeated serial repetitions of the blossom algorithm is limited by restriction to a small window of decoding for each repetition, resulting in a constant-time algorithm.

We can compare the results obtained in Ref.\ \onlinecite{Fow13} to our results by extracting the improvement of correlated over basic fault-tolerant corrections for a distance-3 code. For a depolarization probability comparable to the one
we use in Fig.~\ref{fig:decay_fig} the improvement is approximately $24\%$. This is similar to the improvement we obtained with the neural network decoder. 

Both the neural network decoder and the improved blossom decoder perform below the optimal maximum-likelihood decoder. Several approaches exist to reach the optimal limit, we mention the incorporation of X--Z correlations via a belief propagation algorithm \cite{Cri17}, and approaches based on renormalization group methods or Monte Carlo methods \cite{Bra14,Duc10,Hut14}.

Bravyi, Suchara, and Vargo \cite{Bra14} reported a density-matrix renormalization group (DMRG) method for exact single-round maximum-likelihood decoding in polynomial time, assuming bit-flip and dephasing noise. Their performance continues to be better than the blossom decoder for multi-round decoding. The method is somewhat limited in the choice of error model; in particular it cannot account for Y-errors.

The Markov-chain Monte Carlo method of Hutter, Wootton, and Loss \cite{Hut14} samples over the set of corrections to approximate the maximum-likelihood decoding via the Metropolis algorithm. This again outperforms the blossom decoder, but it suffers from an increased run-time cost, with an additional $O(N^2)$ computational cost.

\subsection{Approaches based on machine learning}

The existence of algorithms \cite{Fow13,Del14,Cri17,Bra14,Duc10,Hut14} that improve on the blossom decoder does not diminish the appeal of machine learning decoders, since these offer a flexibility to different types of topological codes that a dedicated decoder lacks. 

Torlai and Melko \cite{Tor16} implemented a machine learning decoder based on a restricted Boltzmann machine, while Varsamopoulos, Criger, and Bertels \cite{Var17} and Krastanov and Jiang \cite{Kra17} used feedforward neural networks. The key distinction with our work is that we use a recurrent neural network, and thereby allow the decoder to detect correlations between arbitrary cycles of stabilizer measurements. 

Refs.\ \onlinecite{Tor16} and \onlinecite{Kra17} were limited to the study of models without circuit-level noise (\textit{i.e.}~ without measurement error between repeated cycles), and so no direct quantitative comparison with the performance of our decoder is possible.

One feedforward neural network in Ref.\ \onlinecite{Var17} was constructed to take the syndrome from 3 cycles as input. While it cannot decode an arbitrary number of cycles, it can account for circuit noise at the 3-cycle level. Over that time frame their performance lies within error bars from that of our recurrent neural network. (The equivalence of the Pauli-frame-update error rate of Ref.\ \onlinecite{Var17} and our parity-bit error rate is discussed in App.\ \ref{errorrate_appendix}.)

\section{Design of the neural network decoder}
\label{sec_neuralnetwork}
\begin{figure*}
\centering{ 
\includegraphics[width=0.8\textwidth]{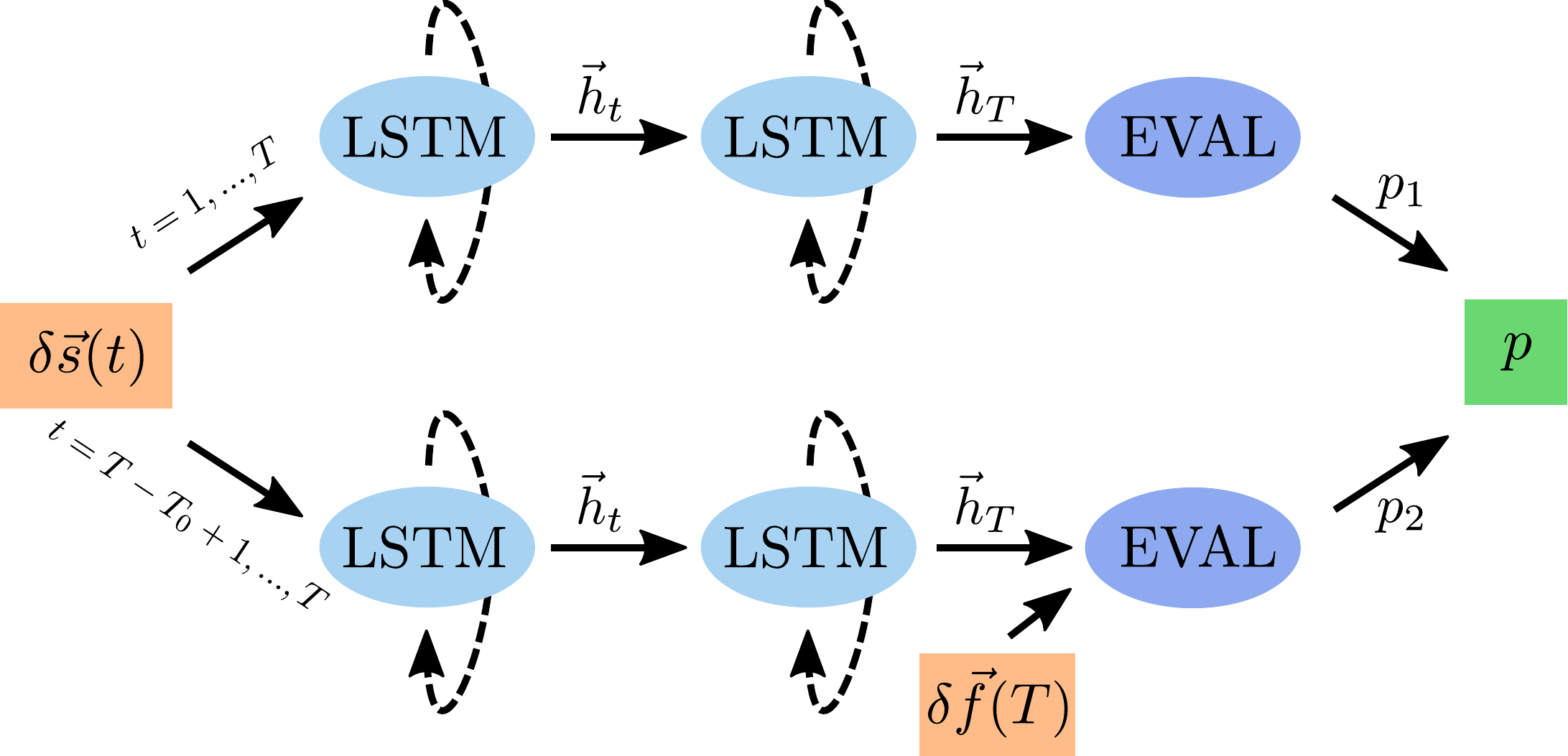}
\caption{\label{fig:network}Architecture of the recurrent neural network decoder, consisting of two neural networks. 
The upper half is network 1 and the lower half is network 2.
Ovals denote the long short-term memory (LSTM) layers and fully connected evaluation layers, while boxes denote input and output data. 
Solid arrows denote data flow in the system, and dashed arrows denote the internal memory flow of the LSTM layers.
}
}
\end{figure*}
The neural network consists of two LSTM layers with internal state sizes $N_L^{(1)}$ and $N_L^{(2)}$, and a fully connected evaluation layer with $N_L^{(E)}$ neurons. We implement the decoder using the \textit{TensorFlow} library \cite{tensorflow}, taking $N_L^{(1)}=N_L^{(2)}=N_L^{(E)}=64$. 
The LSTM layers receive as input sets of syndrome increments $\delta\vec{s}(t)$ from both the $x$-stabilizer and the $z$-stabilizer measurements.

When a final parity prediction is required from the network at time $T$, information from the recurrent network is passed to an evaluation layer, along with the syndrome increment 
\begin{equation}
\delta\vec{f}(T)=\vec{f}-\vec{s}(T)\mod 2\label{eq:final_derivative}
\end{equation}
between final syndrome $\vec f$ calculated from the data qubit measurements and the last syndrome readout $\vec{s}(T)$ from the ancilla qubits. Note that, while $\vec{s}(t)$ is passed to each decoder in both the $x$-basis and the $z$-basis, the final syndrome $\vec{f}$ is only available to a decoder in its own basis.

The memory of the recurrent network solves the issue of how to concatenate multiple decoding cycles, but one remaining issue occurs at the end of the computation: the final syndrome breaks time-translational invariance.
Within any cycle, the decoder must account for the possibility that an error signal ($\delta s_i(t)=1$) should be propagated forward in time to future cycles.
This is not the case for the final syndrome, as this is calculated directly from the data qubit measurements, and any errors in the data qubits do not propagate forward in time.

To achieve time-translational invariance of the decoder we split the problem into two separate tasks, as shown in Fig.~\ref{fig:network}. 
Task 1 is to estimate the probability $p_1$ that the parity of bit-flip errors during T cycles is odd, based solely on the syndrome increments $\delta\vec{s}(t)$ up to that point (\textit{i.e.}~those extracted from ancilla measurements).
Task 2 is to estimate the probability $p_2$ that the final data qubit measurements make any adjustment to the final parity measurement, based solely on new information from the final syndrome increment $\delta\vec{f}(T)$.
The final parity probability is then given by the probabilistic sum
\begin{equation}
 p = p_1 (1-p_2) + p_2 (1-p_1).\label{eq:prob_sum}
\end{equation}

We use two separate networks for the two tasks. The first network gets $T$ rounds of syndrome increments $\delta\vec{s}(t)$ but not the final syndrome increment (upper half of Fig.~\ref{fig:network}). The second network gets only the last $T_0$ syndrome increments $\delta\vec{s}(t)$, but its evaluation layer gets the last output of the second LSTM layer concatenated with the final syndrome increment (lower half of Fig.~\ref{fig:network}). For Surface-$17$, we observe optimal performance when we allow the task-2 network a window of $T_0=3$ cycles, giving a decoder that works for experiments of three or more cycles. In general, the number of cycles fed to the second network should be on the order of the length of the longest time-correlations between syndromes. As task 2 only requires decoding of a fixed number of cycles, it could potentially be performed by a simpler feedforward network, but we found it convenient to keep the same architecture as task 1 because of the similarity between the two tasks. 

We discuss the details of the network architecture and training procedure in App.\ \ref{details_appendix}. The source code is available \cite{Decoder}.

\section{Neural network performance}
\label{sec_performance}
\begin{figure}
\includegraphics[width=\columnwidth]{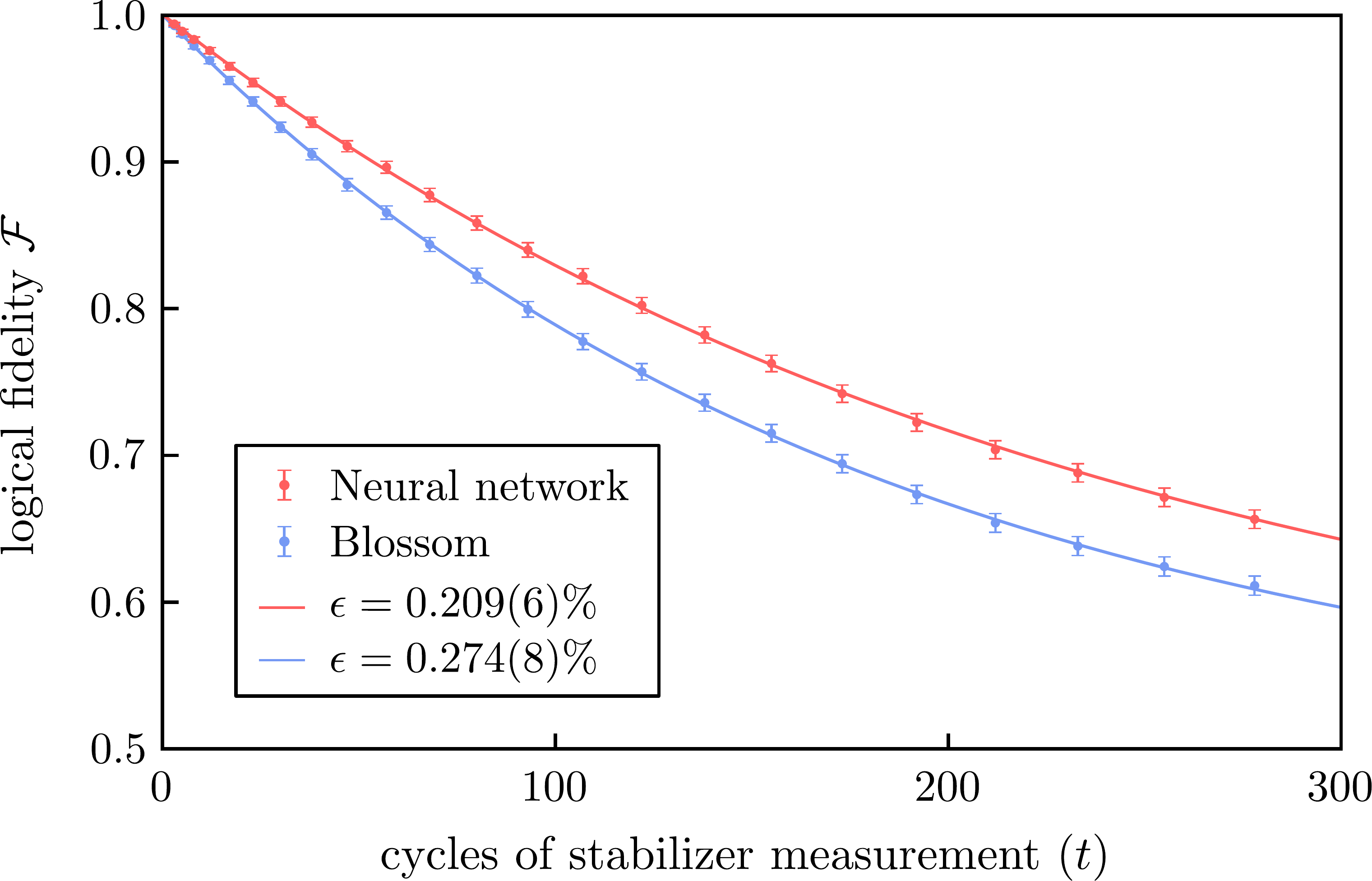}
\caption{\label{fig:decay_fig}Comparison of logical qubit decay between blossom and neural network decoders for a Pauli error channel model, with $p_x=p_y=p_z= 0.048\%$ and $p_m= 0.14\%$.
We plot the probability that the decoder corrects the logical qubit after $t$ cycles of stabilizer measurement and error accumulation.
All data is averaged over $5\cdot 10^4$ datasets, with error bars obtained by boot-strapping (using $3\sigma$ for the error).
Lines are two-parameter fits of the data to Eq.~\eqref{eq:fidelity}.}
\end{figure}
We determine the neural network performance on the 17-qubit distance-$3$ surface code, referred to as ``Surface-$17$'', which is under experimental development~\cite{Ver16}.

We take at first a simplified Pauli error channel model~\cite{CircuitModel}, similar to Refs.\ \onlinecite{Fow12b, Tom14} but without correlated two-qubit errors. In this model the performance of the blossom decoder is understood and individual error types can be focused upon.
Stabilizer measurements are made by entangling two or four data qubits with an ancilla qubit, which is readout in the computational basis (right panel in Fig.~\ref{fig:Surface_code}).

The process is broken into seven steps: four coherent steps over which {\sc cnot} gates are performed, two steps in which Hadamard gates are performed, and one measurement step. 
During idle, Hadamard, and {\sc cnot} steps, both data and ancilla qubits have independent chances of a $\sigma_x$ error (with probability $p_x$), a $\sigma_y$ error (with probability $p_y$), and a $\sigma_z$ error (with probability $p_z$). This implies that the total probability during any step for a qubit to accumulate a $y$-error (as opposed to an $x$-error, a $z$-error, or no error) is
\begin{equation}
\mbox{$y$-error prob.}=p_y(1-p_x)(1-p_z) + p_xp_z(1-p_y).
\end{equation}
With this definition $p_y=0$ implies that $x$-errors and $z$-errors are uncorrelated (it does not imply that there are no $y$-errors).

Data qubits behave similarly during measurement steps, but ancilla qubits are projected into the computational basis and so cannot incur phase errors.
Instead, a measurement has a $p_m$ chance of returning the wrong result, without the qubit state being affected.
Qubits are reused after measurement without reset, and so the syndromes $s_i(t)$ are obtained by changes in the readout $m_i(t)$ of an ancilla qubit between rounds,
\begin{equation}
s_i(t)=m_i(t)-m_i(t-1)\mod 2.
\end{equation}

The performance of the logical qubit is measured using the protocol outlined in Ref.\ \onlinecite{Obr17} (Methods section).
The logical qubit is prepared in the $|0\rangle$ state, held for $T$ cycles, and finally measured and decoded.
The decoder seeks to determine whether or not the qubit underwent a logical bit-flip during this time.
The probability that the decoder obtains the correct answer gives the logical qubit fidelity, which can be plotted as a function of the number of cycles.
Fig.~\ref{fig:decay_fig} shows the decay in fidelity over 300 cycles for $p_x=p_y=p_z= 0.048\%$ and $p_m= 0.14\%$, which corresponds to a physical error rate of approximately $1\%$ per cycle.

A logical error rate per cycle $\epsilon$ can be obtained from these figures by a two-parameter fit to the logical fidelity
\begin{equation}
\mathcal{F}(t)=\tfrac{1}{2}+\tfrac{1}{2}(1-2\epsilon)^{t-t_0},\label{eq:fidelity}
\end{equation}
where $t_0$ is a constant offset to account for the `majority vote' behavior of the error correcting circuit at low cycle number~\cite{Obr17}, and any additional sample preparation and measurement error.
We find $\epsilon=0.209\%$ for the neural network decoder, a substantial improvement over the value $\epsilon=0.274\%$ for the blossom decoder~\cite{MWPM}.

\begin{figure}
\includegraphics[width=\columnwidth]{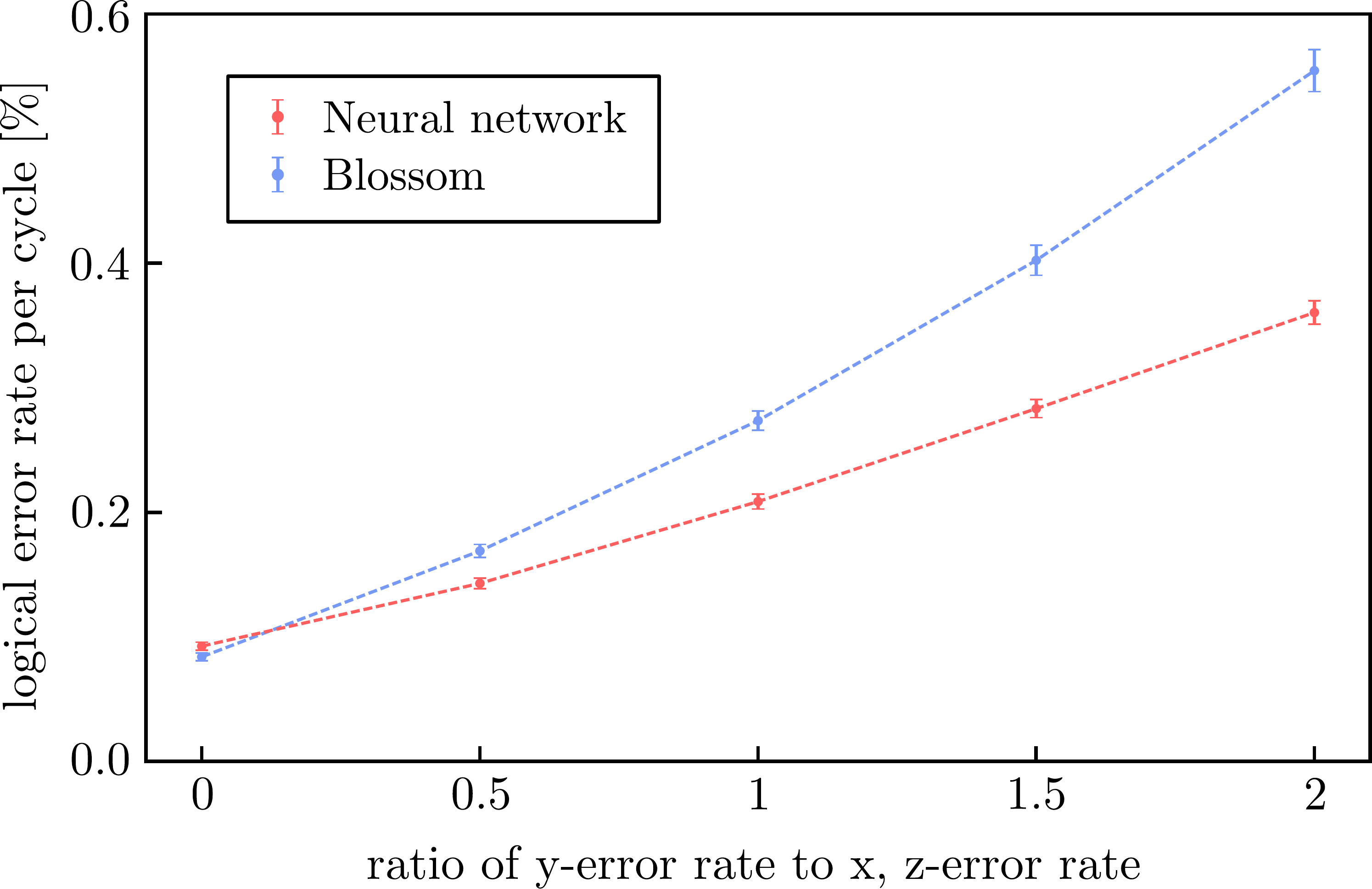}
\caption{\label{fig:scaling_fig}Comparison of the error rates $\epsilon$ of a logical qubit decoded by a neural network and a blossom decoder, for different values of the correlated error rate $p_y$. As $p_y$ increases, at fixed $p_x=p_z= 0.048\%$ and $p_m= 0.14\%$, the blossom decoder (blue) produces a larger error rate than the neural network decoder (red). Data points are obtained by fitting decay curves, as in Fig.\ \ref{fig:decay_fig}.
}
\end{figure}

To demonstrate that the performance improvement is due to the capability of the neural network to detect error correlations, we show in Fig.\ \ref{fig:scaling_fig} how the performance varies with varying $p_y$ (at fixed $p_x=p_z= 0.048\%$ and $p_m= 0.14\%$).
When $p_y=0$, the $\sigma_x$ and $\sigma_z$ errors are independent and the blossom decoder performs near-optimally~\cite{Obr17,Hei16}. The neural network decoder then gives no improvement, but once $p_y\sim p_x$ the performance gain is evident.

\section{Conclusion and outlook}
\label{sec_conclusion}
\begin{figure}
\includegraphics[width=\columnwidth]{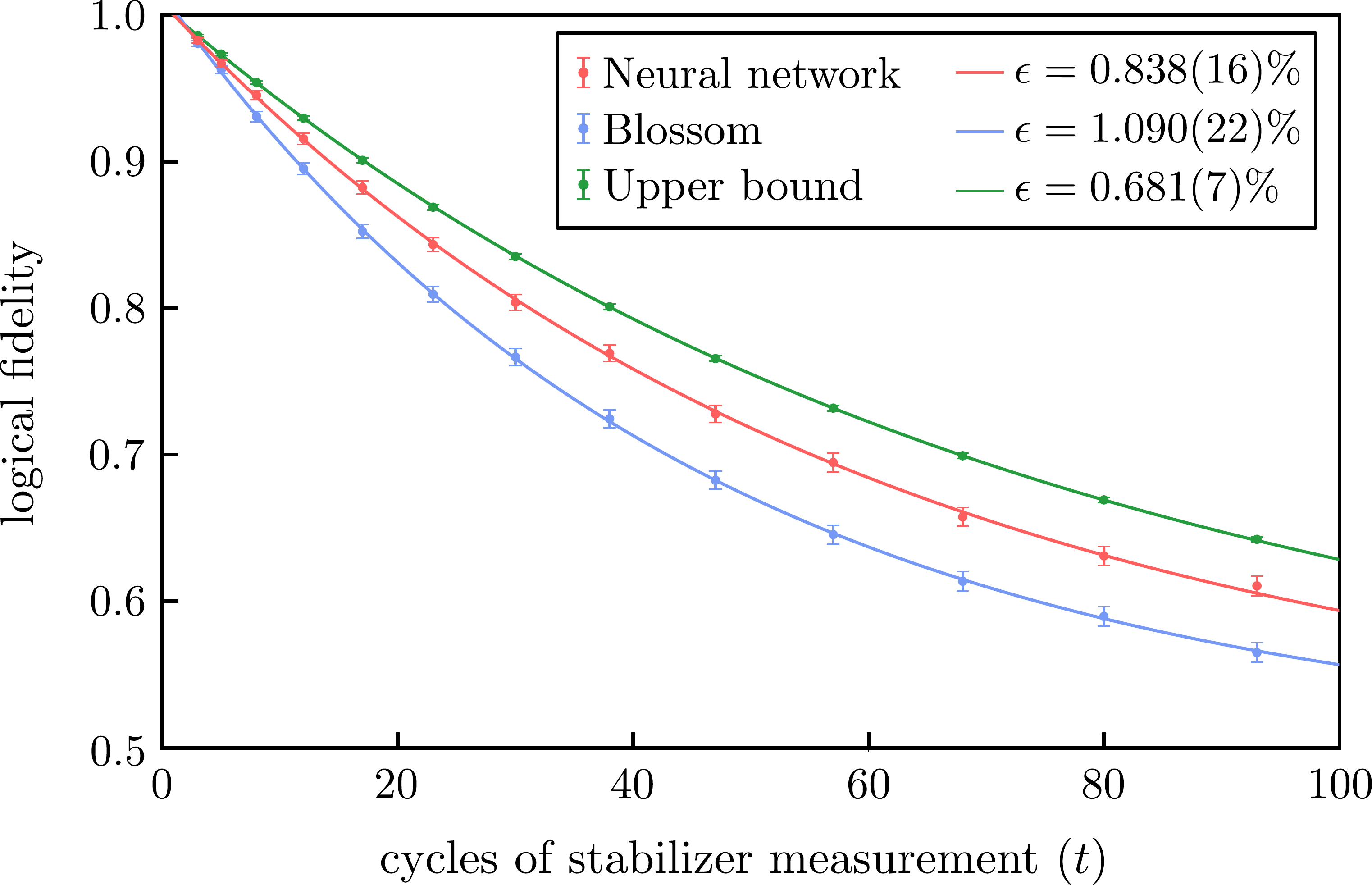}
\caption{\label{fig:bm_qs} Same as Fig.\ \ref{fig:decay_fig}, but now for a density matrix simulation of an implementation of Surface-17 using superconducting transmon qubits \cite{Obr17}.
}
\end{figure}
In conclusion, we have designed and tested a recurrent neural network decoder that outperforms the standard minimum-weight perfect matching (MWPM, or ``blossom'') decoder in the presence of correlated bit-flip and phase-flip errors.
The building block of the network, a long short-term memory layer, allows the decoder to operate over the full duration of a quantum algorithm with multiple cycles.
A key feature of our design, which sets it apart from independent proposals \cite{Var17,Kra17,note0}, is that the network can be trained solely on experimental data, without requiring \textit{a priori} assumptions from theoretical modeling.

We believe that our neural network decoder provides a realistic option for utilization in forthcoming experimental QEC implementations \cite{Ver16}. In support of this, we have tested the performance in a real-world setting by using a density matrix simulator to model Surface-$17$ with state-of-the-art experimental parameters for superconducting transmon qubits~\cite{Obr17}. In Fig.~\ref{fig:bm_qs} we show the decay of the fidelity over 100 cycles for the neural network and blossom decoders, as well as an upper bound on the optimal fidelity. (The latter is extracted directly from the simulation data.) The decoder efficiency \eqref{eq:decoder_efficiency} of the neural network is $\eta_d=0.81$, a $26\%$ improvement over the blossom decoder.
This improvement was achieved after training on $4 \cdot 10^6$ datasets, which require roughly $60\,{\rm s}$ to generate on experimental hardware~\cite{Ver16}, making this approach immediately experimentally viable.

We mention three directions for future research. The first is the extension to other topological codes than the surface code, such as the color code. The neural network itself is agnostic to the type of topological code used, so this extension should be feasible without modifications of the design.
Secondly, for low error rates it will be challenging to train a neural network decoder, because then the training dataset is unlikely to contain a sufficient representation of two-qubit errors.
This can potentially be overcome by training on data with a higher error rate, but it remains to be seen whether a decoder trained this way will outperform MWPM decoding.
Finally, the decoder needs to be scaled-up to surface codes that are deformed by lattice surgery~\cite{Hor12} or braiding~\cite{Bra98} for the execution of logical gates. For this extension the design of the decoder should be modified so that it is not tied to a single code distance.

\acknowledgments
We have benefited from discussions with B. Criger, L. DiCarlo, A. G. Fowler, V. Ostroukh, and B. Terhal.
This research is supported by the Netherlands Organization for Scientific Research (NWO/OCW), an ERC Synergy Grant, and by the Office of the Director of National Intelligence (ODNI), Intelligence Advanced Research Projects Activity (IARPA), via the U.S. Army Research Office grant W911NF-16-1-0071. The views and conclusions contained herein are those of the authors and should not be interpreted as necessarily representing the official policies or endorsements, either expressed or implied, of the ODNI, IARPA, or the U.S. Government. The U.S. Government is authorized to reproduce and distribute reprints for Governmental purposes notwithstanding any copyright annotation thereon.

\appendix

\section{Details of the neural network decoder}
\label{details_appendix}

\subsection{Architecture}
The decoder is composed of two networks. The first network maps a list of syndrome increments $\delta \vec s(t)$ with $t=1, 2, ..., T$ to a probability $p_1\in [0,1]$.
The second network maps a list with the last few syndrome increments $t=T-T_0+1, T-T_0+2, ..., T$, together with a single final syndrome increment $\delta \vec f(T)$ to a probability $p_2\in [0,1]$.
The probabilistic sum $p=p_1(1-p_2)+p_2(1-p_1)$ of these two outputs is the probability that the logical qubit has incurred a bit-flip error.
The cost function we try to minimize is the cross-entropy between this probability and the true final parity of bit-flip errors (labels) plus a small weight regularization term.

We note that $p$ is invariant under the transformation $p_1\rightarrow 1-p_1$ and $p_2\rightarrow 1-p_2$.
This ambiguity in the individual error probabilities is irrelevant for the joint operation of the networks. Moreover, it may be easily removed by testing the trained networks separately on a trivial example where all syndromes are zero and both probabilities should be $<1/2$.

Both networks consist of two LSTM layers with internal states $\vec c_t^i, \vec h_t^i \in \mathbb R^{64}$ and a fully connected evaluation layer with 64 rectified linear units.
The inputs of the first layer are the syndrome increments. The inputs of the second layer are the outputs of the first layer $\vec h_t^1$. For the first network, the input of the evaluation layer is the final output of the second LSTM layer, subject to a rectified linear activation function $\textnormal{ReL}(\vec h_T^2)$.
For the second network, the input of the evaluation layer is $\textnormal{ReL}(\vec h_T^2)$ concatenated with the final syndrome increment $\delta \vec f(T)$.

The source code including all the network parameters is available~\cite{Decoder}.

\subsection{Training and evaluation}
The two networks are trained simultaneously on mini-batches of size 64 from a training dataset containing $4\cdot 10^6$ sequences of lengths between $T=11$ and $T=20$ cycles. At the end of each sequence, the training set contains the final syndrome increment and the final parity of bit-flip errors. One epoch consists of $10^4$ mini-batches. The optimization is done using the Adam optimizer \cite{Kin14} with a learning rate of $10^{-3}$. For regularization we apply dropout \cite{Hin12} with a keep probability of $0.8$ after each LSTM layer and after the evaluation layer. In addition, we apply weight decay with a prefactor of $10^{-5}$ to the evaluation layer. After each epoch, the decoder is evaluated on a validation dataset, which consists of $10^4$ sequences of lengths between $T=81$ and $T=100$ cycles. If the logical error rate on the validation dataset reaches a new minimum, the network is stored. The training continues until the logical error rate on the validation dataset has not improved for $100$ epochs. We train three decoders and choose the instance that has the lowest logical error rate on the validation dataset.

To evaluate the chosen decoder, we use yet another dataset. This test dataset consists of $5\cdot 10^4$ sequences of length $T=300$ for the Pauli error channel model and $T=100$ for the density matrix simulation. In contrast to the training and validation datasets, the test dataset contains a final syndrome increment and a final parity of bit-flip errors after each cycle. This cannot be achieved in a real experiment, but is extracted from the simulation to keep the calculation time manageable. We evaluate the decoder on the test dataset for $t_n=2+\sum_{n'=1}^n n' \leq T$ cycles, chosen such that the resolution is high at small cycle numbers and lower at large cycle numbers. If the decoders output is $p<0.5$, the final parity of bit-flip errors is predicted to be even and otherwise odd. We then compare this to the true final parity and average over the test dataset to obtain the logical fidelity. Using a two-parameter fit to Eq.\ \eqref{eq:fidelity} we obtain the logical error rate per cycle.

\section{Parity-bit error versus Pauli-frame-update error}\label{app:PauliFrame}
\label{errorrate_appendix}

Ref.\ \onlinecite{Var17} described the error rate of the decoder in terms of its ability to apply the correct Pauli frame update. The error rate $\epsilon$ from Eq.\ \eqref{eq:fidelity} describes the correctness of the parity bit produced by the decoder, without explicitly referring to a Pauli frame update. Here we show that the two error rates are in fact the same.

We recall that a Pauli frame is a list of Pauli X, Y, or Z errors that have occurred to data qubits  \cite{Ter15}. Two Pauli frames are equivalent if they are separated by stabilizer measurements, since these act as the identity on the error-free subspace. 

We begin by choosing logical operators $X_L$ and $Z_L$ in terms of Pauli operators on the physical qubits. The choice is not unique because of a gauge freedom: $SX_L= X_L$ on the logical subspace for any stabilizer operator $S$. 

Consider a syndrome $\vec{s}(t)$ that contains only a single non-zero stabilizer measurement $s_i(t)$, corresponding to a stabilizer operator $S_i$. There exist multiple Pauli frames $P_i$ that correct $S_i$ and which commute with our chosen logical operators. Ref.\ \onlinecite{Var17} considers a `simple' decoder, which arbitrarily chooses one $P_i$ for each $S_i$. Then, given a syndrome $\vec{s}(t)$ at time $t$ with many non-zero $s_i$, it generates a Pauli frame as
$P_{\mathrm{simple}}=\prod_{i,s_i(t)=1}P_i$.

The simple decoder is coupled to a neural network decoder, which outputs a parity bit $p$ that 
determines whether or not to multiply $P_{\mathrm{simple}}$ by $X_L$ (if the neural network is calculating Z-parity) or $Z_L$ (if the neural network is calculating X-parity). We denote the resulting Pauli frame update by $P_{\rm calc}$. If it differs from the true Pauli frame update $P_{\rm true}$ the decoder has made an error, and the rate at which this happens is the Pauli frame update error rate $\epsilon_P$.

To see that this $\epsilon_P$ is equivalent to the parity-bit error rate $\epsilon$, we consider for the sake of definiteness a neural network that calculates the Z-parity. The two Pauli frames $P_{\mathrm{calc}}$ and  $P_{\mathrm{true}}$ differ by $X_L$ when $[P_{\mathrm{true}},Z_L]\neq[P_{\mathrm{calc}},Z_L]$. But $[P_{\mathrm{true}},Z_L]$ is the parity readout of the data qubits, and $[P_{\mathrm{calc}},Z_L]$ is precisely our prediction. Alternatively, note that the simple decoder is constructed to fix $[P_{\mathrm{simple}},Z_L]=0$, and the choice to multiply this by $X_L$ precisely fixes $[P_{\mathrm{calc}},Z_L]=p$.

We finally note that in a physical experiment the Pauli frame $P_{\mathrm{true}}$ is undetermined unless the data qubits themselves are measured in the Z or X basis, and the gauge freedom is fixed at random by this measurement. The parity bit $p$ is therefore not only more convenient for a neural network to output than a Pauli frame update, but also more appropriate, as this way the neural network does not spend time trying to predict the outcome of quantum randomness.

\end{document}